\documentclass[prl,aps, twocolumn, showpacs,groupedaddress,10pt,nofootinbib]{revtex4-1}

\usepackage[T1]{fontenc}
\usepackage{bm,dcolumn}
\usepackage{amsmath,amsfonts,amssymb,amsthm,amscd} 
\usepackage{mathrsfs}

\usepackage{graphicx}
\usepackage{verbatim}
\usepackage[section]{placeins}
\usepackage{color}
\usepackage{soul}
\usepackage{xparse}
\usepackage{physics}
\usepackage{siunitx}

\graphicspath{ {./Figures/} }
\usepackage{pdfpages} 
\usepackage{pgffor}

\makeatletter
\AtBeginDocument{\let\LS@rot\@undefined}
\makeatother

\begin{document}

\title{A Wavefunction Microscope for Ultracold Atoms}

\author{S. Subhankar}
\thanks{These two authors contributed equally}
\affiliation{Joint Quantum Institute, National Institute of Standards and Technology and the University of Maryland, College Park, Maryland 20742 USA}

\author{Y. Wang\normalfont\textsuperscript{*,}}
\email{Corresponding author. wang.yang.phy@gmail.com}

\affiliation{Joint Quantum Institute, National Institute of Standards and Technology and the University of Maryland, College Park, Maryland 20742 USA}

\author{T-C. Tsui}
\affiliation{Joint Quantum Institute, National Institute of Standards and Technology and the University of Maryland, College Park, Maryland 20742 USA}

\author{S. L. Rolston}
\affiliation{Joint Quantum Institute, National Institute of Standards and Technology and the University of Maryland, College Park, Maryland 20742 USA}

\author{J. V. Porto}
\affiliation{Joint Quantum Institute, National Institute of Standards and Technology and the University of Maryland, College Park, Maryland 20742 USA}

\date{\today}

\begin{abstract}

Quantum simulations with ultracold atoms typically create atomic wavefunctions with structures at optical length scales, where direct imaging suffers from the diffraction limit. In analogy to advances in optical microscopy for biological applications, we use a non-linear atomic response to surpass the diffraction limit.  Exploiting quantum interference, we demonstrate imaging with super-resolution of 
$\lambda/100$ and excellent temporal resolution of 500 ns. We characterize our microscope's performance by measuring the ensemble averaged wavefunction of atoms
within the unit cells of an optical lattice, and observe the dynamics of atoms excited into periodic motion. This approach can be readily applied to image any atomic or molecular system, as long as it hosts a three-level system. 

\end{abstract}

\pacs{37.10.Jk, 32.80.Qk, 37.10.Vz}

\maketitle

High spatial and temporal resolution microscopy can reveal the underlying physics, chemistry, and biology of a  variety of systems. Examples range from the study of atoms on surfaces with atomic-resolution scanning tunneling microscopy (STM)\cite{Binnig1986} to the use of super-resolution microscopy to observe individual molecule dynamics within living cells \cite{Hell2007}. The field of quantum simulation with ultracold atoms has emerged to study strongly correlated manybody systems using the precise control with light-atom interactions \cite{Gross2017}. This entails confining atoms, engineering their interactions and potentials, and measuring them with laser light. Based on fluorescence and absorption, the inherent imaging resolution is limited by diffraction. Bringing super-resolution microscopy to the field of quantum simulation of condensed matter systems will allow new direct probes of the wavefunction in a  variety of systems that simulate, for example, many-body localization \cite{Schreiber842}, periodically driven superconductors \cite{Mitrano2016}, high temperature superconductivity \cite{Lee2006}, and topological insulators \cite{Qi2011}.

We demonstrate here an approach \cite{Miles2013} for imaging atoms with unprecedented spatial resolution $\sim$5.7 nm that is well below the diffraction limit. This allows us 
to directly measure the wavefunction optically within the unit cells of a 1D optical lattice, in contrast to measuring site occupancies \cite{Gross2017, Andrea2016, Nelson2007}. Far-field microscopy at the nanoscale based on nonlinear optical response is well established \cite{Hell2007} to resolve molecular dynamics inside biological samples. Using similar ideas, subwavelength addressing \cite{Gorshkov2008} and localized excitation has been proposed \cite{Paspalakis2001, Chang2006,Yavuz2007, Li2008, Yang2018, Ashida2015} and observed \cite{Miles2013,Maurer2010}. Based on the dark state associated with a three-level system \cite{Gorshkov2008,Paspalakis2001,Chang2006,Yavuz2007,Li2008,Yang2018,Miles2013}, we coherently shelve narrow slices of the wavefunction in every unit cell of the lattice into one of the spin states dictated by the local dark state. We selectively read out the total population in that spin state, which is proportional to the local probability density of the lattice wavefunction. The working resolution (width of the slice) can be adjusted with the dark state composition, and is ultimately limited by the signal-to-noise ratio (SNR). The coherent nature of this approach allows us to measure on a timescale much faster than the evolution of the wavefunction. Our setup can be readily applied to current quantum gas experiments. By dispersively coupling the readout state to a cavity, as suggested by Ref \cite{Yang2018}, we could perform  subwavelength QND measurements.

\begin{figure}[!t]
\includegraphics[width=8.cm]{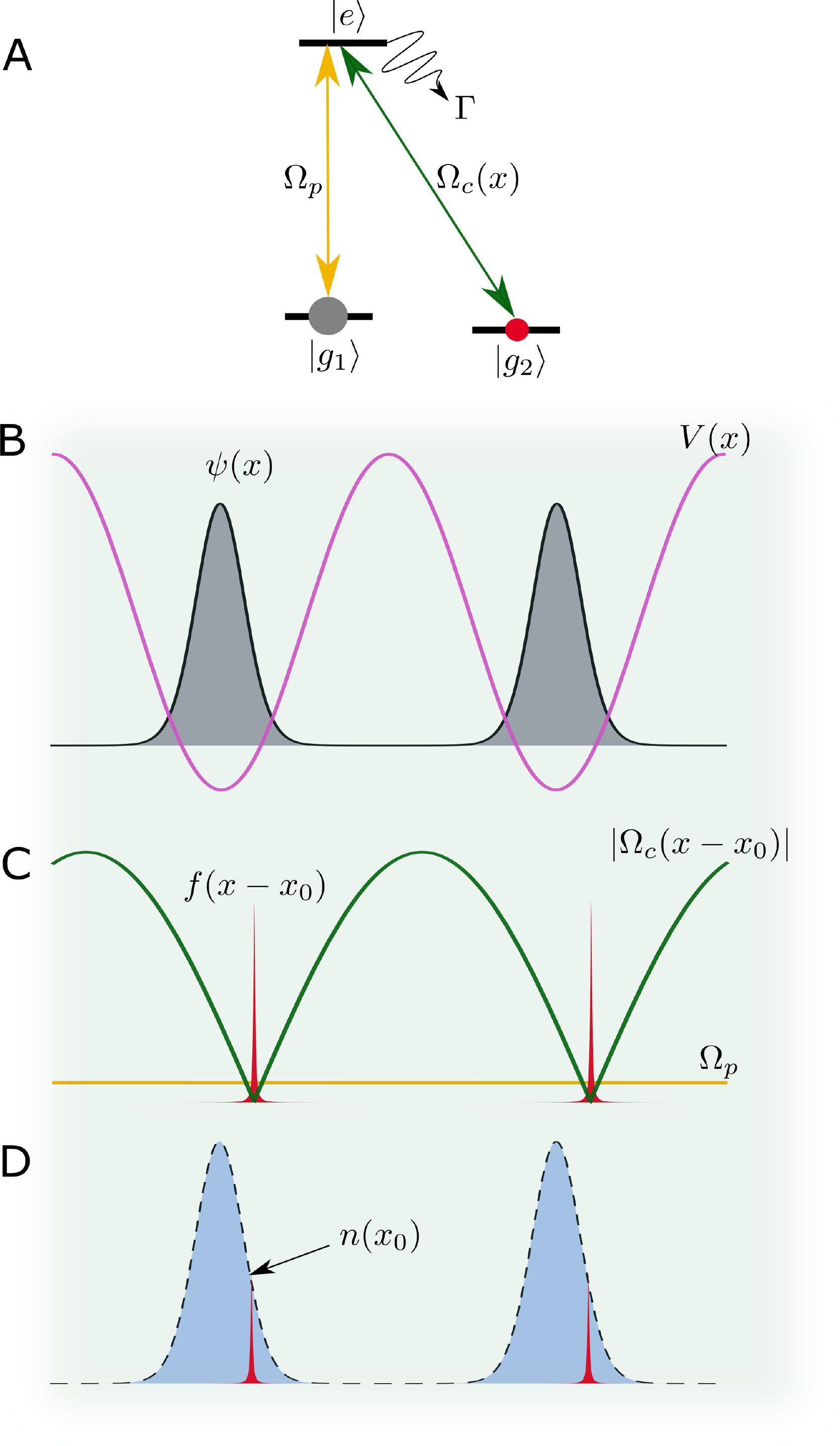}
\caption{ Principle of a wavefunction microscope. 
\textbf{(A)} Configuration of the control field $\Omega_c(x)$ and probe field $\Omega_p$.
\textbf{(B)} Wavefunction $\psi(x)$ in $|g_1\rangle$  in the lattice of interest $V(x)$. 
\textbf{(C)} The spin state composition is transferred to $|g_2\rangle$ near the nodes of $\Omega_c(x-x_0)$ with probability density given by $f(x-x_0)$ (narrow red peaks), and $|g_1\rangle$ elsewhere. The width of $f(x-x_0)$ is determined by the relative strength of the two light fields $\epsilon=\Omega_p/\Omega_c$. \textbf{(D)} $f(x-x_0)$ maps $|\psi(x)|^2$ onto the population in $|g_2\rangle$, $n(x_0)$, which can be selectively measured via state-dependent imaging. By stepping through different positions $x_0$ and measuring $n(x_0)$, we can reconstruct $|\psi(x)|^2$.
}
\label{fig:schematic}
\end{figure}

The principle of our approach is illustrated in Fig.1 and is similar to Refs \cite{Miles2013,Yang2018,Yang2018a}. Assuming adiabaticity, a three-level atom (Fig.1A) coupled by two spatially-varying light fields will stay in a dark state, which is decoupled from the excited state. This dark state is a superposition of the two ground states with spatially-varying amplitudes:
\begin{equation}
|D(x)\rangle =\frac{1}{\sqrt{\Omega_c(x)^2+\Omega_p^2}} ( \Omega_c(x) |g_1\rangle - \Omega_p |g_2\rangle).
\end{equation}
Here, we use a standing wave control field $\Omega_c(x)=\Omega_c \sin(k x)$ and homogeneous probe field $\Omega_p$, where $k=2\pi/\lambda$, and $\lambda$ is the wavelength of the light. The resulting dark state composition is predominantly $|g_1 \rangle$ away from the nodes of $\Omega_c (x)$, and $|g_2\rangle$ near the nodes where $\Omega_p\gg|\Omega_c(x)|$. The probability density of $|g_2\rangle$ (Fig.1C), coming from this nonlinear dependence on the Rabi frequencies (Eq.1), is periodic and has narrow peaks near the nodes:
\begin{equation}
f(x)=\frac{\epsilon^2}{\epsilon^2+\sin^2(kx)}
\end{equation}
where $\epsilon=\Omega_p/\Omega_c$. The half-width-at-half-maximum (HWHM) $\sigma$ of the peaks provides a good metric for the resolution within the unit cell $\lambda/2$. For small $\epsilon$, $\sigma$ depends linearly on $\epsilon$: $\sigma \simeq \epsilon\lambda/2\pi$. We would ideally expect $\sigma\simeq 2\textrm{ nm} \simeq \lambda/280$ for $\epsilon=0.02 $. Much like an STM, the wavefunction probability density $|\psi(x)|^2$ (Fig.1B) can be determined by measuring population in $|g_2\rangle$ at different locations $x_0$ (Fig.1D), yielding a signal
\begin{equation}
n(x_0)=\int |\psi(x)|^2f (x-x_0)dx.
\end{equation}
By deconvolving this signal with the probing function $f(x)$, we can reconstruct $|\psi(x)|^2$.

We use stimulated Raman adiabatic passage (STIRAP) \cite{Vitanov2017} to transfer the selected slices of the wavefunction from the state $|g_1\rangle$ into $|g_2\rangle$. In order to accurately measure the shape of the wavefunction, the STIRAP process must be adiabatic with respect to the spin degree of freedom (i.e., the dark state composition given by Eq. 1), but diabatic with respect to the motional degree of freedom. For small $\epsilon$, the shortest duration of the STIRAP is inversely proportional to the Rabi frequencies  \cite{Supplementary}. For typical trapped atoms experiments, Rabi frequencies can be tens of MHz, while the motional dynamics is on the order of tens of kHz. 

We work with the three-level system in $^{171}$Yb consisting of $|g_1\rangle= |^1S _0, F=\frac{1}{2}, m_F=-\frac{1}{2} \rangle$,  $|g_2\rangle=|^1S_0, F=\frac{1}{2}, m_F=+\frac{1}{2}\rangle$,  and $|e\rangle=|^3P_1, F=\frac{1}{2}, m_F=-\frac{1}{2}\rangle$. The control field $\Omega_c(x)$ is formed by two counter-propagating $\sigma^-$-polarized beams $\Omega_{c1}e^{ikx}$ and $\Omega_{c2}e^{-ikx}$ in the direction of the quantization axis defined by a magnetic field along $\hat{x}$, while the probe field $\Omega_p$ is a $\pi-$polarized traveling wave normal to the control beams \cite{wang2017}. We prepare $^{171}$Yb atoms by sympathetically cooling them with $^{87}$Rb atoms \cite{Vaidya2015}. After ramping up the magnetic field to 36 mT and removing the Rb atoms, the Yb atoms are optically pumped into $|g_1\rangle$ with a final population N $\sim 2\times 10^5$. We measure the wavefunction of spin-polarized Yb atoms loaded into either a Kronig-Penney (KP) type lattice of thin barriers, as described in Ref \cite{wang2017}, or a regular sinusoidal lattice based on the ac Stark shift of $\Omega_{c1,2}$ off-resonantly coupled to the $|g_1\rangle \leftrightarrow |^3P_1, F=\frac{3}{2}, m_F=-\frac{3}{2}\rangle$ transition \cite{Supplementary}. 

Our microscope is implemented as follows. We first suddenly turn off the lattice potential $V(x)$ that supports the wavefunction  to be probed by switching off the $\Omega_{c2}$ beam. 
Next, we ramp on $\Omega_p$ followed by $\Omega_{c2}$ with a different phase, which adiabatically flips the spin from $|g_1\rangle$ to $|g_2\rangle$ in the region tightly localized near the nodes of the shifted $\Omega_c(x-x_0)=\Omega_c\sin(k(x-x_0))$. The intensity profiles for ramping these two beams are calculated to preserve adiabaticity, ensuring atoms follow the spatio-temporal dark state at all times. 
We then rapidly ramp off all beams simultaneously in order to preserve the dark state composition. We measure the $|g_2\rangle$ population via state-selective absorption imaging after time-of-flight. Scanning $x_0$ in fine steps at small $\epsilon$ allows us to map out the wavefunction with high resolution. 

\begin{figure}[!h]
\includegraphics[width=8.cm]{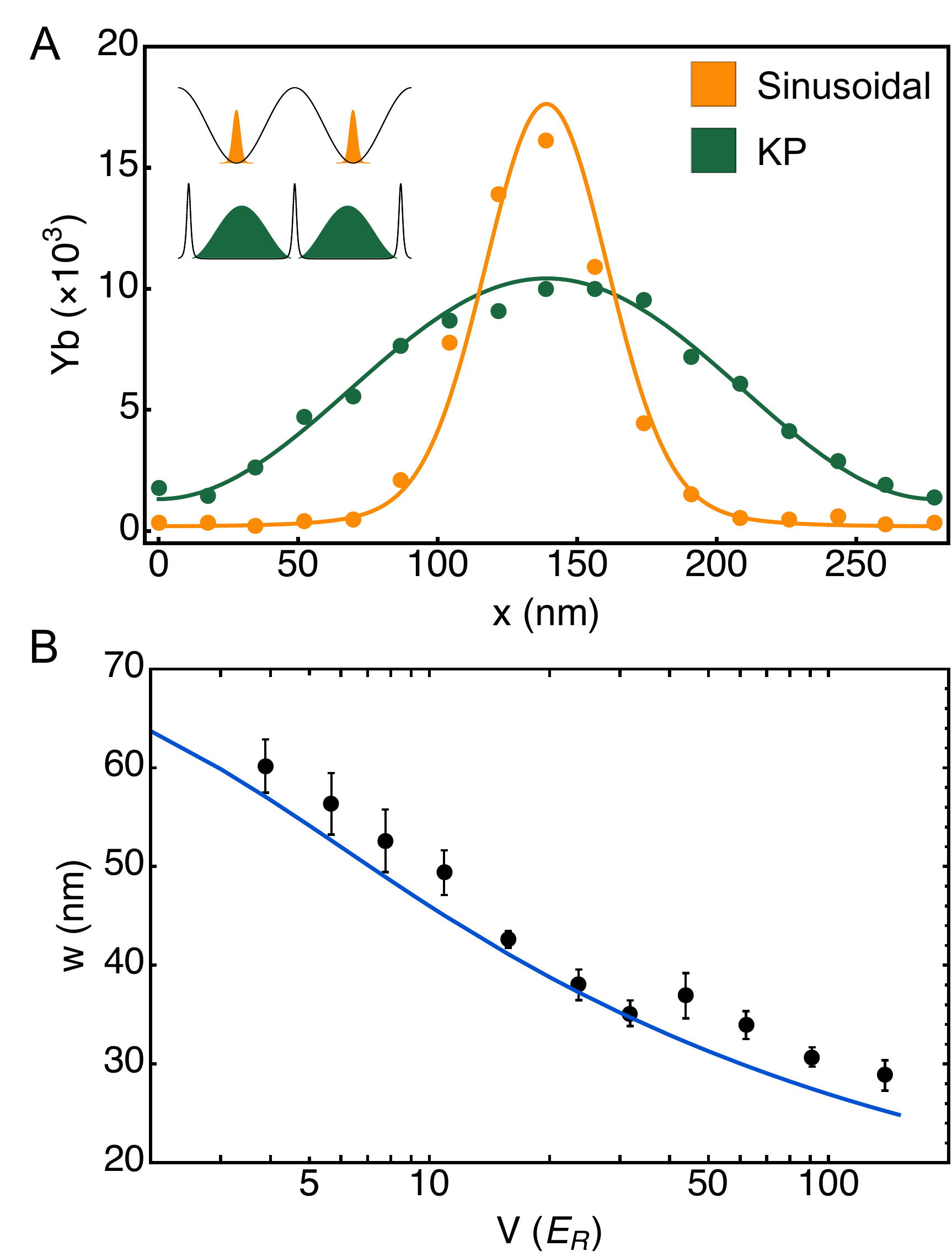}
\caption{Measurements of the ground state wavefunction within the unit cell of an optical lattice with different shapes.
\textbf{(A)} The orange points show $n(x)$ for atoms in a sinusoidal lattice, measured with $\epsilon=0.05$. The green points represent $n(x)$ in a Kronig-Penney lattice, measured with $\epsilon=0.1$. 
The colored lines are calculations normalized to the same atom number.
Inset: schematic of different lattice potentials and corresponding wavefunctions.
\textbf{(B)} $w$ of $n(x)$ in a sinusoidal lattice as a function of the lattice depth. Black points show experimental data with $\epsilon=0.05$, and the blue line is a calculation including the 800 ns measurement time. The error bars are one standard deviation from the Gaussian fits. 
}
\label{fig:staticlattices}
\end{figure}

We use our wavefunction microscope to investigate atoms in sinusoidal and KP lattices. We start by preparing the atoms in the ground band of the lattice of interest \cite{Supplementary}. 
Fig. \ref{fig:staticlattices}A shows $n(x)$ measured in a 140 $E_R$ sinusoidal lattice using a theoretical resolution of 4.4 nm,  along with $n(x)$ in a KP lattice with 50 $E_R$ barriers using a theoretical resolution of 8.8 nm. Here, $E_R=\hbar^2k^2/2m$ is the recoil energy, and $m$ is the mass of the atom.
The different lattice potential (sinusoidal vs. box-like KP) gives rise to different functional forms of the wavefunction in the lattice (Inset of Fig.\ref{fig:staticlattices}A). The expected wavefunction is Gaussian for a deep sinusoidal lattice, and cosine for the KP lattice. The solid lines are the calculated functional forms including the resolution.
In Fig. \ref{fig:staticlattices}B we show the HWHM, $w$, of the ground-band wavefunction of the sinusoidal lattice as a function of lattice depth. The blue curve represents the calculated width, 
taking into account the wavefunction expansion during the 800~ns total measurement time. The prediction is in good agreement with the data. The remaining discrepancy may result from trap inhomogeneities, the uncertainty of the Rabi frequencies, and mechanical effects arising from the non-adiabatic potentials due to the spatially-varying dark state \cite{wang2017,Jendrzejewski2016,Lacki2016}.

\begin{figure}[!t]
\includegraphics[width=8.cm]{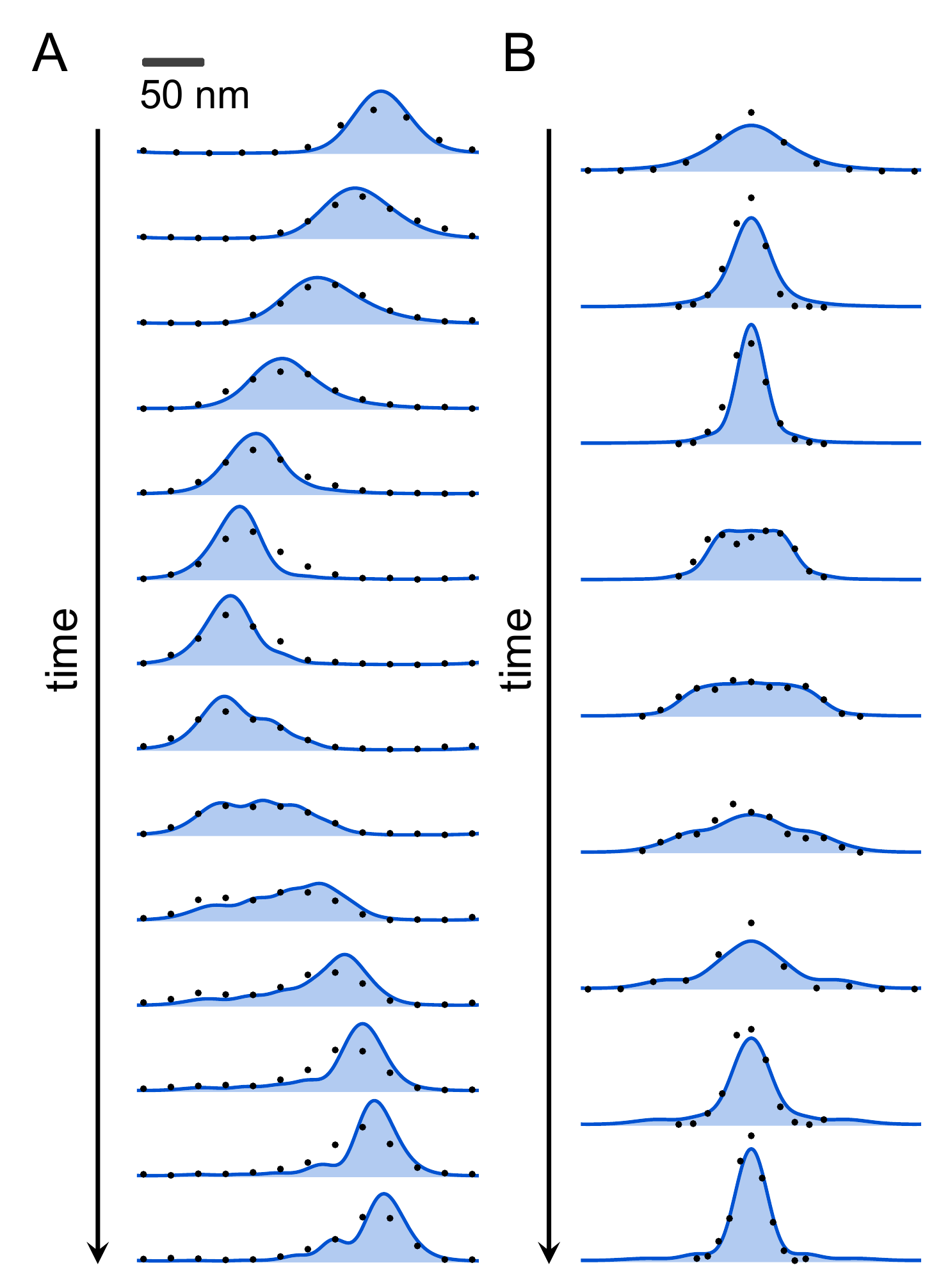}
\caption{ Wavefunction dynamics within the unit cell of an optical lattice. We excite 
\textbf{(A)} sloshing motion, and 
\textbf{(B)} breathing motion of $\psi(x)$ in a 140 $E_R$ deep sinusoidal lattice by suddenly changing either the position or the depth of the lattice potential. $n(x)$ is plotted at different hold times (1 $\mu$s to 14 $\mu$s in steps of 1 $\mu$s for A, and 1.5 $\mu$s to 9.5 $\mu$s in steps of 1 $\mu$s for B) after the sudden change. The points are experimental data with $\epsilon$ = 0.05 and the blue curves represent calculations of $n(x)$ based on the independently measured lattice parameters. 
} 
\label{fig:dynamics}
\end{figure}

The fast time scale for the STIRAP slicing process allows for observing wavefunction dynamics. At our maximum Rabi frequency of $\Omega_c=2\pi\times90$ MHz and $\epsilon=0.05$, we can maintain the adiabaticity condition for a STIRAP time of 500 ns. Fig. \ref{fig:dynamics}A shows the dynamics of the wavefunction in a sinusoidal lattice after a sudden shift in the lattice position. The atoms are first adiabatically loaded into a 140 $E_R$ lattice. Then the lattice position is diabatically changed in 100 ns by $\lambda/8$ via the phase of the $\Omega_{c1}$ lattice beam, which excites ``sloshing'' motion. We map out the temporal and spatial evolution of the wavefunction within the unit cell by holding the atoms in the shifted lattice for incremental periods of time before probing. In Fig \ref{fig:dynamics}A, we show the measured wavefunctions, which are in  agreement with the calculations.

The dynamics of the wavefunction after a sudden change in lattice depth is shown in Fig. \ref{fig:dynamics}B. The atoms are first  adiabatically loaded into the ground-band of a shallow lattice (6 $E_R$). The depth of the lattice is then suddenly increased to 140 $E_R$, which excites ``breathing'' motion of atoms inside a unit cell. As time increases, we see the  wavefunction breathes at a frequency characterized by the band energies. At $t=3.5$ $\mu$s and $t=9.5$ $\mu$s, the wavefunction is focused to $w \sim$ 13 nm.

\begin{figure}[!t]
\includegraphics[width=8.cm]{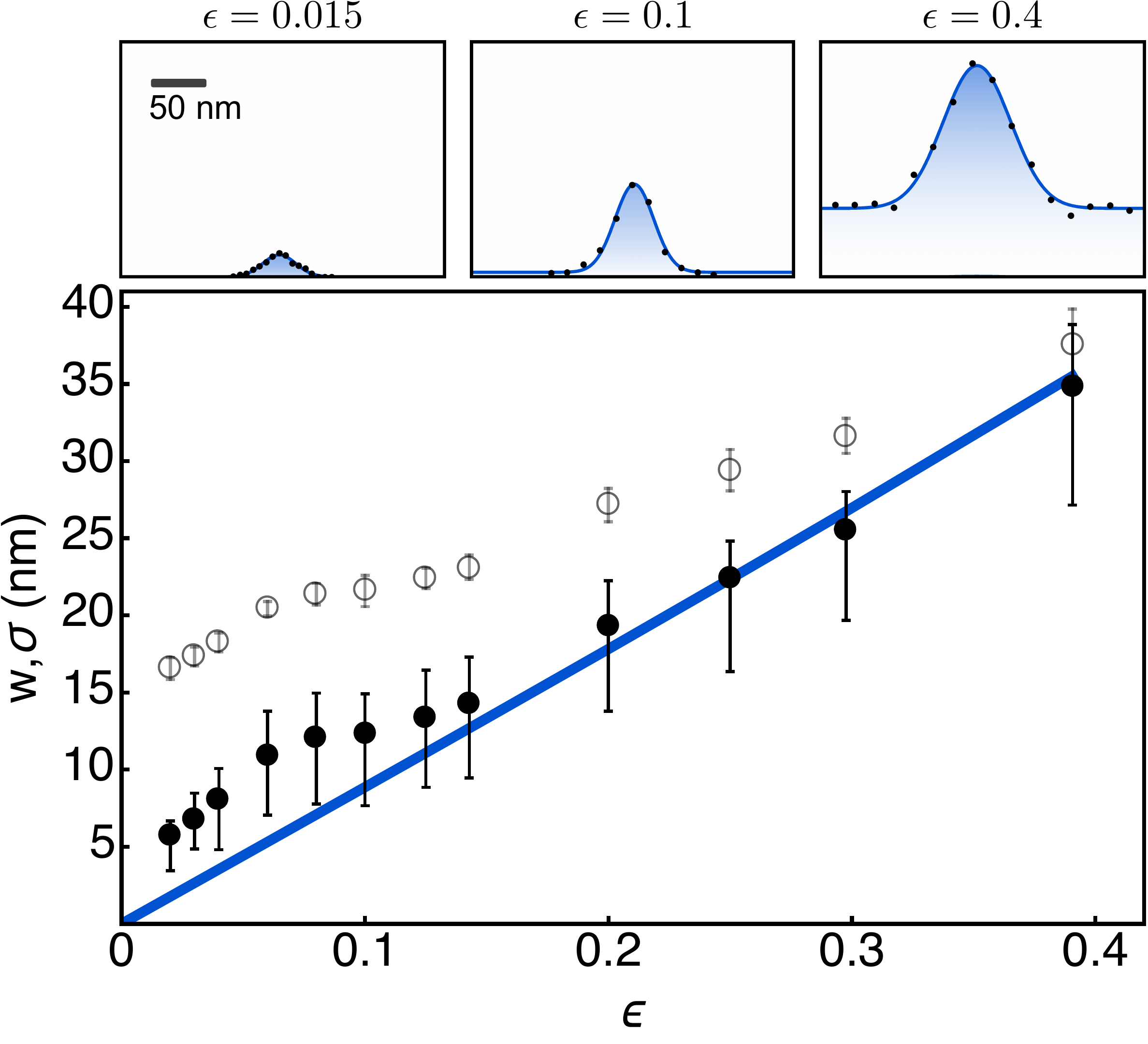}
\caption{ Spatial resolution of the microscope. 
We create narrow wavefunction $|\psi(x)|^2$ ($w\sim$ 13 nm) by exciting the breathing motion of atoms in a deep sinusoidal lattice and measure $n(x)$ at the focus point (see Fig. \ref{fig:dynamics} B) as a function of $\epsilon$. The measured $w$ from a Gaussian fit with an offset to the $n(x)$ (see upper panel for typical wavefunction measurements) is plotted against $\epsilon$ as the gray open circles, with error bar showing one standard deviation from the fitting. These data are then deconvolved with the calculated wavefunction $|\psi(x)|^2$ to find the intrinsic resolution $\sigma$ and plotted as the black closed circles. 
The error bars are dominated by the systematic uncertainties in $w$. 
The blue curve is the calculated resolution of $f(x)$ at different $\epsilon$.
} 
\label{fig:resolution}
\end{figure}

We estimate the spatial resolution of our microscope by measuring the narrowest wavefunction that we create after the breathing mode excitation. This occurs at 9.5 $\mu$s where the theoretically expected wavefunction has $w$=$13.1^{+0.8}_{-0.3}$ nm, where the uncertainty arises from the uncertainty in the Rabi frequency calibrations.
We measure $n(x)$ at this stage with different resolution by varying $\epsilon$, as shown in Fig.\ref{fig:resolution}. The measured width $w$ of $n(x)$ is plotted as the gray open circles, which decrease and approach the expected value for small $\epsilon$. 
By deconvolving the results with the calculated wavefunction $|\psi(x)|^2$, we determine the intrinsic resolution $\sigma$ for different $\epsilon$, which is plotted as the black points. The inferred resolution agrees with the calculated resolution (blue solid line). The ultimate resolution is possibly limited by mechanical effects arising from the sharp potential associated with the dark state \cite{wang2017,Jendrzejewski2016,Lacki2016}. 
As the slice width $\sigma$ decreases, the total population in $|g_2\rangle$ also decreases, setting a practical limit on the usable resolution, 
as illustrated by wavefunction measurements shown in Fig \ref{fig:resolution} upper panel. The smallest measured $\sigma$ reaches $5.7^{+1.0}_{-2.2}$~nm, which could be improved with higher SNR and Rabi frequencies.

In conclusion, we have demonstrated super-resolution imaging of wavefunctions in optical lattices with a spatial resolution of $\lambda/100$ and a temporal resolution of 500 ns. This imaging technique (demonstrated here on an ensemble of atoms) can be extended to single atoms by averaging over multiple realizations. This dark-state based technique can be applied to image any atomic or molecular system as long as they  host a three-level system, 
including the alkali atoms that are used in many experiments.  Such high spatial and temporal resolution microscopy provides a new tool to address ultracold atom simulations of condensed-matter  systems.  For example, periodic driving (Floquet physics) system has been shown to produce superconductivity \cite{Mitrano2016} as well as topological insulators \cite{Wang2013}.  This Floquet dressing is well suited to cold atom simulation and the temporal resolution of our microscope will allow a window into the dynamic evolution of the wavefunction during the periodic cycle.  Finally, while the imaging technique demonstrated here measures the wavefunction probability density, the coherence of the dark-state selection process could allow for measurement of the local wavefunction phase as well.

\paragraph*{Note added} Recently, we become aware of similar work from Cheng Chin's group.  

\appendix
 \bibliographystyle{apsrev4-1.bst}

\section*{Acknowledgments}
We thank Victor M. Galitski, Alexey V. Gorshkov and Przemyslaw Bienias for fruitful discussions. This work is supported by NSF PFC at JQI and ONR (Grant No. N000141712411) 



\section*{Supplementary material}

# Supplementary Material: A Wavefunction Microscope for Cold Atoms

S. Subhankar,[*] Y. Wang,[*] T-C. Tsui, S. L. Rolston, and J. V. Porto
*Joint Quantum Institute, National Institute of Standards and Technology
and the University of Maryland, College Park, Maryland 20742 USA*
(Dated: July 8, 2018)

## I. $^{171}$YB ATOM LEVEL STRUCTURE

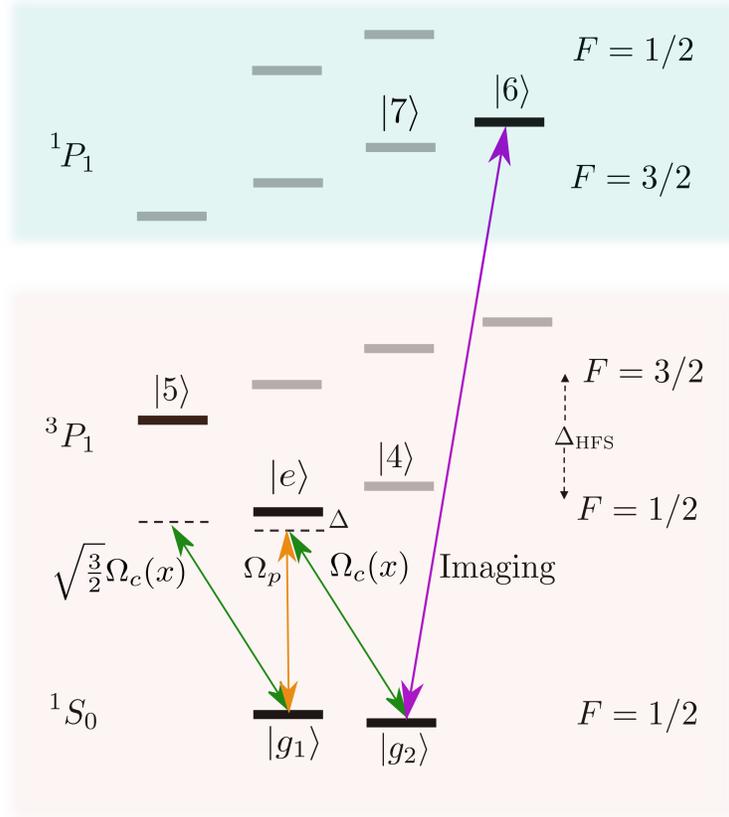

FIG. S1: Level structure of the $^1S_0$, $^3P_1$, and $^1P_1$ manifolds of $^{171}$Yb: $\Delta$ is the single photon detuning; and $\Delta_{\text{HFS}} \sim 6$ GHz is the $^3P_1$ hyperfine splitting.

We measure the Yb atoms by absorption imaging on the $^1S_0 - {}^1P_1$ transition with light at 399 nm generated by a frequency doubled laser system. We stabilize the seed of the imaging laser (at 800 nm) via a scanning transfer cavity lock [S1, S2] with the master laser locked to a saturated absorption feature on the $^{85}$Rb $|5^2S_{1/2}, F = 3\rangle \leftrightarrow |5^2P_{3/2}, F' = 3-4\rangle$ crossover signal. State-selectivity is achieved by imaging in a large magnetic field of 36 mT along $\hat{x}$, such that the resulting 440 MHz separation between $|6\rangle$ and $|7\rangle$ Zeeman sublevels of $^1P_1$ is much larger than the linewidth $\Gamma_{^1P_1} = 2\pi \times 27.9$ MHz. The imaging beam propagates along $\hat{x}$ with $\sigma^+$-polarization relative to $\vec{B}$. We measure the population in the $|g_2\rangle$ hyperfine ground state by making the laser resonant with its respective stretched state, $|g_2\rangle \leftrightarrow |6\rangle$ for 10 $\mu$s. Optical pumping of atoms in $|g_1\rangle$ into $|g_2\rangle$ via $|7\rangle$ is suppressed by a factor of over 3000 due to the 440 MHz detuning and the short imaging time.

---

[*] These two authors contributed equally

Fig. S1 depicts the three hyperfine states that constitute the $\Lambda$-system, consisting of $|g_1\rangle, |g_2\rangle$, and $|e\rangle$, that we use to generate the KP lattice and to probe the wavefunction of arbitrary lattices. We create the off-resonant sinusoidal ac-Stark Shift lattices using the $|g_1\rangle \leftrightarrow |5\rangle$ transition with the lattice depth given by $\frac{3\Omega_c^2(x)}{8\Delta_{HFS}}$. The effect of this off-resonant lattice is negligible when the atoms are in the KP lattice with $\Omega_c = 70\Gamma$ and $\Omega_p = 10\Gamma$. The method we use to calibrate our Rabi frequencies is detailed in Ref. [S3].

## II.  EXPERIMENTAL SEQUENCE

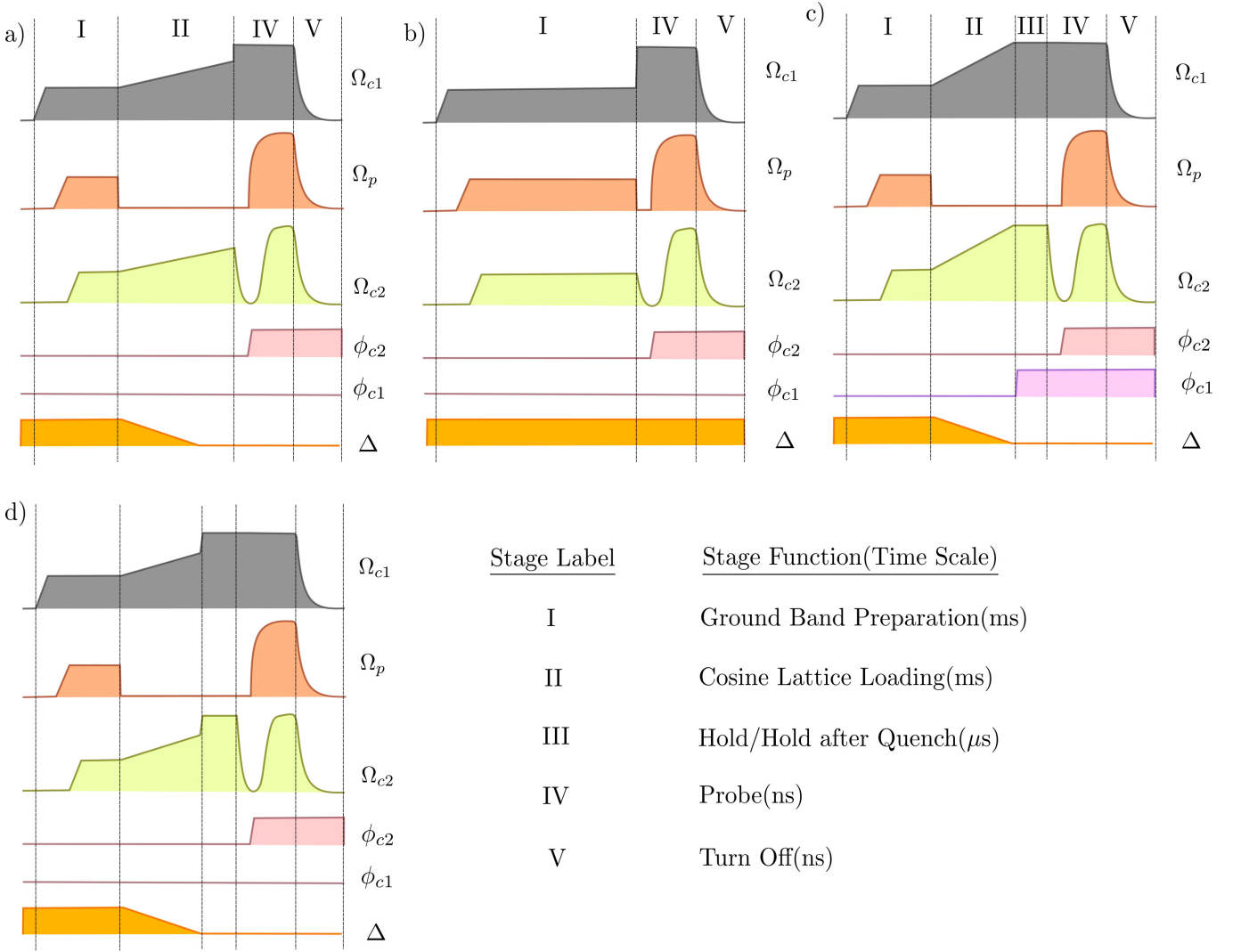

FIG. S2: Experimental sequences. a) Probing the ground state wavefunction of a sinusoidal lattice. b) Probing the ground state wavefunction of a KP lattice. c) Probing the dynamics after a sudden change in the lattice position. d) Probing the dynamics after a sudden change in lattice depth.

Preparation and experimental sequence: Before the start of each experimental sequence, the atoms are optically pumped into $|g_1\rangle$.

*Stage I:* To simplify the study of the static and dynamics properties of wavefunctions in lattices, we prepare our atom cloud to fill only the ground band of the lattice of interest. Since the Fermi energy of our atomic cloud is $\sim 3E_R$, adiabatic loading into the lattice will have some population in the first excited band. We resolve this issue by first loading atoms into a KP lattice with $\epsilon = 0.14$ ($\Omega_{c1} = \Omega_{c2} = 35\Gamma$, $\Omega_p = 10\Gamma$ and $\Delta = 4$MHz)[S3] and then holding for 5 ms. Atoms in higher bands of the KP lattice have a shorter lifetime and are lost from the trap, effectively removing atoms in the higher bands.

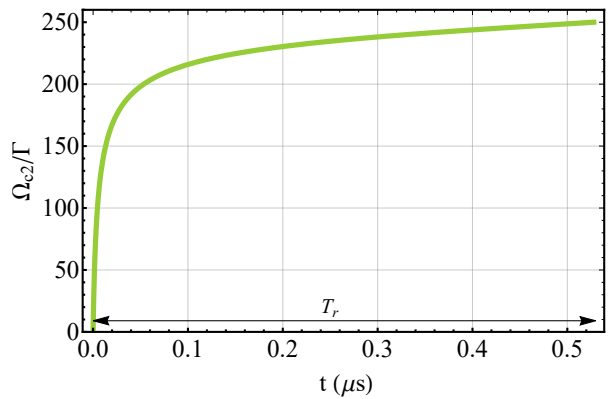

FIG. S3: The optimal amplitude waveform for $\Omega_{c2}(t)$ for $\Omega_{c1} = 250\Gamma$, $\Omega_p = 25\Gamma$

*Stage II:* In this stage we adiabatically transfer atoms from the ground band of the KP lattice into the ground band of an ac-Stark shift lattice in 10 ms. $\Delta$ is ramped down to 0, which is important in achieving the maximum speed while adiabatically following the dynamic dark state in Stage IV.

*Stage III:* In this stage we excite dynamics in the lattice. In Fig.S2c, the phase of the $\Omega_{c1}$ beam is ramped to 90° in 100 ns, so as to diabatically shift the position of the lattice by $\lambda/8$ which induces sloshing dynamics. In Fig.S2d, the lattice depth is suddenly increased from $6E_R$ to $140E_R$ and the atoms are held in the deep lattice for different times(1.5 $\mu$s - 9.5 $\mu$s) to study the breathing motion.

*Stage IV:* In this stage we measure the wavefunction. First, $\Omega_{c2}$ is suddenly turned off to 0 while $\Omega_{c1}$ is set to $250\Gamma$. Then $\Omega_p$ beam is suddenly turned on to its desired value. Due to the large energy separation of $125\Gamma$ ($\Omega_{c1} = 250\Gamma$ and $\Omega_p = \Omega_{c2} = 0$) between the dark and bright states, the adiabatic following of the dark state is guaranteed during the turn-on of the $\Omega_p$ beam. Then $\Omega_{c2}$ beam is turned on with a different phase, $\phi_{c2}$, implemented by changing the phase of the RF drive to the AOM, with the amplitude being ramped up to $250\Gamma$ with the optimal waveform so as to preserve the adiabaticity during the ramp. By scanning $\phi_{c2}$ from 0 to 360°, we change the position of the node of $\Omega_c(x) = 500\Gamma \cos(kx)$, thereby mapping out the probability amplitude of atoms in each spatial slice of the wavefunction.

*Stage V:* Finally, the lattice beams are ramped off simultaneously in 100 ns by switching off the RF drive to the AOMs. Since the dark-state composition only depends on the ratio $(\Omega_{c1}+\Omega_{c2})/\Omega_p$ and not on the absolute magnitude of the Rabi frequencies, simultaneous ramp-off of the lattice beams preserves the dark state composition until the atoms are imaged.

### III. HARDWARE CONTROL

In order to generate the experimental sequences described earlier, we need to have fine, high bandwidth control over the amplitude and phase of the light fields $\Omega_{c1}$, $\Omega_{c2}$, and $\Omega_p$. This control is achieved by using a home-built FPGA(Spartan 6)-controlled DDS(AD9910) based RF signal generator. We use three such devices to drive three acousto-optic modulator(AOMs) for the light fields. Phase coherence between the light fields is ensured by having the devices be clocked by the same 10 MHz clock source and having the light fields be derived from the same laser. Each device generates an 80 MHz RF carrier signal with arbitrary amplitude and phase, and imprints that onto the light via its respective AOM. The DDS can update the phase of the RF signal every 4 ns. The desired amplitude waveform (AW)is loaded into the local RAM of the FPGA of the device and is updated at a maximum update rate of 8 ns. The maximum length of AW pulse is $\sim 256$ $\mu$s when updated every 8 ns.

### IV. OPTIMAL AMPLITUDE WAVEFORM FOR STIRAP

During stage IV, we adiabatically transfer atoms from $|1\rangle$ to $|2\rangle$ near the node of $\Omega_c(x)$ via Stimulated Raman adiabatic passage (STIRAP) [S4]. For an ideal $\Lambda$-system, the local adiabatic criterion is given by Ref. [S4] as $\Omega_{\rm rms} \gg |\Omega_c\dot{\Omega}_p - \dot{\Omega}_c\Omega_p|/\Omega_{\rm rms}^2$, where $\Omega_{\rm rms} = \sqrt{\Omega_c^2 + \Omega_p^2}$ (at $\Delta = 0$) is the energy gap between the dark and bright

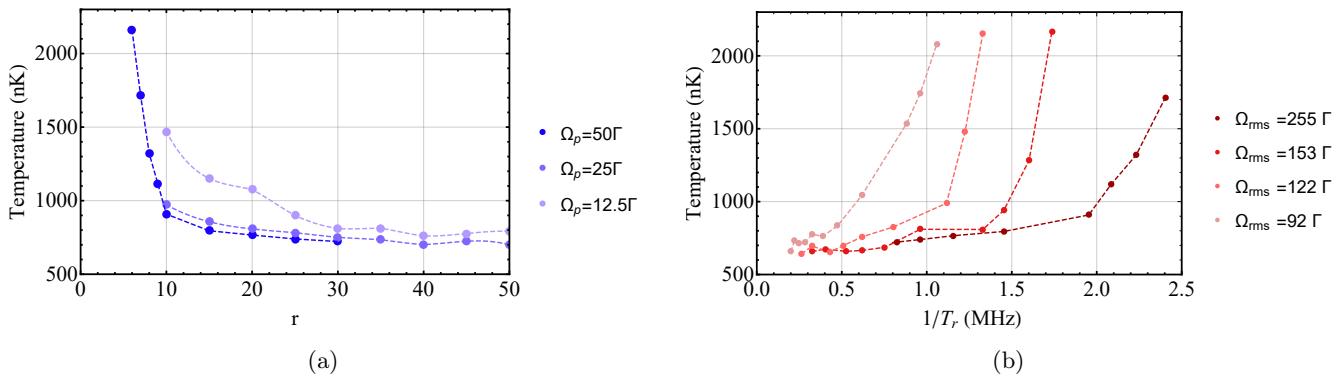

FIG. S4: a)The temperature of the atoms after 10 complete STIRAP pulses for a) different values of $r$ and $\Omega_p$, and b) different values of $\Omega_{\rm rms}$ and $T_r$

eigenstates and the RHS is the off-diagonal coupling between them. We define an adiabaticity parameter $r$,

$$\Omega_{\rm rms} = r \frac{|\Omega_c \dot{\Omega}_p - \dot{\Omega}_c \Omega_p|}{\Omega_{\rm rms}^2}. \tag{S1}$$

A larger value of $r$ implies a more adiabatic, but slower transfer. The equation is solved to give an optimal shape of $\Omega_{c2}$ near the node of $\Omega_c(x)$ ($\Omega_{c1}$ and $\Omega_p$ are kept constant here) for stage IV:

$$\Omega_{c2}(t) = \Omega_{c1} - \Omega_p \frac{\left(\frac{\Omega_{c1}}{\sqrt{\Omega_{c1}^2 + \Omega_p^2}} - \frac{\Omega_p t}{r}\right)}{\sqrt{1 - \left(\frac{\Omega_{c1}}{\sqrt{\Omega_{c1}^2 + \Omega_p^2}} - \frac{\Omega_p t}{r}\right)^2}}. \tag{S2}$$

The time it takes to finish the $\Omega_{c2}(t)$ ramp is:

$$T_r = \frac{r}{\Omega_p} \frac{1}{\sqrt{1 + 4\epsilon^2}} \tag{S3}$$

For a typical value of $r = 15$, $\Omega_{c1} = 250\ \Gamma$ and $\Omega_p = 25\ \Gamma$, $T_r$ is 0.52 $\mu s$, see Fig. S3. For a given spatial resolution determined by $\epsilon$, more available laser power will reduce $T_r$ and increase the temporal resolution.

We experimentally investigated the minimum $r$ required to ensure adiabatic following of the dynamic dark state. We do this by keeping $r$ fixed and measuring the temperature of the cloud after STIRAP pulses. If the adiabaticity is not well satisfied, the non-zero probability of atoms being in the excited state, $|e\rangle$, leads to scattering which increases the temperature of the atoms. To increase the sensitivity of the measurement, we apply 10 successive STIRAP pulses. This study is performed with only one control beam $\Omega_{c2}$ and the probe beam $\Omega_p$. In each pulse $\Omega_{c2}$ is ramped up from 0 to $250\Gamma$ and then ramped down to 0 following the optimal waveform described by Eq. S1. After 10 cycles, the temperature of the cloud is measured and the results are shown in Fig. S4a. One can see that beyond a certain value of $r$, the STIRAP process becomes adiabatic, i.e., the temperature is independent of $r$. This is about $r = 15$ for $\Omega_p = 50\Gamma, 25\Gamma$. Below $r = 15$, the local adiabaticity criterion breaks down. During the probe stage in Fig. S2, we use $r = 15$ in Eq. S2 to calculate the optimal AWs.

The energy gap between the dark and bright eigenstates increases with increasing $\Omega_{\rm rms}$, which reduces $T_r$ needed to ensure adiabaticity. We study this by keeping the ratio of $\Omega_p/\Omega_{c2} = 0.2$ constant while $\Omega_{\rm rms}$ is varied. As shown in Fig. S4b, with larger $\Omega_{\rm rms}$, faster ramp speed can be achieved while still being adiabatic.

## V.  PRESERVING THE DARK STATE COMPOSITION DURING RAMP-OFF

The ramp-off stage of the lattice beams is crucial for our measurement. This is to preserve the dark-state composition generated during the probing stage. We achieve this by ramping down the light fields simultaneously while maintaining

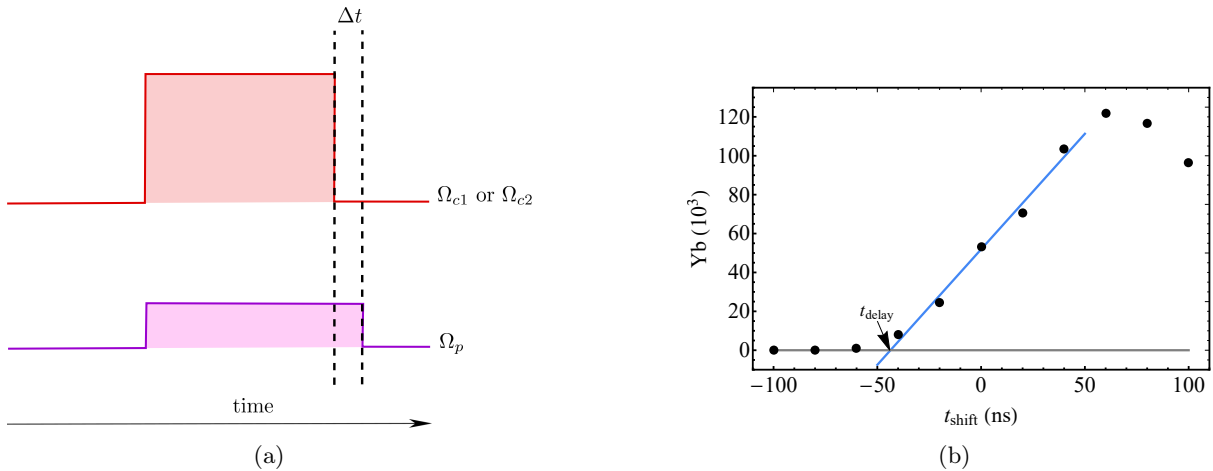

FIG. S5: a)Pulse sequence used to optimize the temporal overlap between the $\Omega_{ci}$ and $\Omega_p$ light fields. b)The population of the atoms in the $|g_2\rangle$ state as a function of $\Delta t = t_{\text{shift}} - t_{\text{delay}}$ with $\Omega_{c1} = 250\Gamma$ and $\Omega_p = 50\Gamma$.

a fixed ratio between the Rabi frequencies $\Omega_{ci}(t)/\Omega_p(t)$, where $i = 1, 2$. The dark-state composition is thus preserved as it is only dependent on the ratio and not on the absolute magnitudes of the Rabi frequencies. For the typical Rabi frequencies we use in the experiments, the relative delay between the light fields needs to be sub-20 ns to preserve the dark state composition. The ramp-off must be diabatic with respect to the mechanical degrees of freedom of the wavefunction. This is guaranteed by turning off the RF drive to the AOMs simultaneously in 100 ns.

Experimentally, simultaneous turn-off of the light fields between different AOMs is not guaranteed as delays may exist due to the laser light hitting the AOM crystals at different distances from their respective piezoelectric transducers. With our best alignment of the AOMs, we reduce the delay of the light fields to within 60 ns of each other. This remnant delay at the atoms is compensated by delaying the TTL signal that turns off the RF drive to an AOM. We accurately measure the turn-off delays between the light fields at the atoms using the method shown in Fig. S5a. Sweeping $t_{\text{shift}}$ of the TTL signal, thereby changing $\Delta t = t_{\text{shift}} - t_{\text{delay}}$, we measure the atom number in $|g_2\rangle$ using state-selective imaging. We are able to change $t_{\text{shift}}$ at ps timescales using a delayed pulse generator(SRS DG535). When $\Delta t < 0$, the spin-composition of the dark-state wavefunction is $|g_1\rangle$. But as $\Delta t \geq 0$, the dark-state composition starts to become predominantly $|g_2\rangle$ with increase in $\Delta t$ as shown in Fig. S4b. By fitting a line to the data, we get the time at which the dark-state spin composition just starts to change from $|g_1\rangle$ to $|g_2\rangle$. The $|g_2\rangle$ component of the dark-state is close to 0 for the Rabi frequencies used in the measurement, when $\Delta t = 0$. By measuring these delays for each pair of beams $\Omega_{c1}\&\Omega_p$, $\Omega_{c2}\&\Omega_p$, we can compensate them via adjusting the lengths of BNC cables of the TTL signals to the RF sources.

Another approach to ensure that the dark state composition does not change is to turn-off the light fields *diabatically* with respect to the spin-degree of freedom of the dark-state wavefunction. As the Rabi frequencies of the light fields are in the range of hundreds of MHz, the turn-off time must be less than 10 ns. It is challenging to achieve such turn-off times with AOMs, but one could use EOMs instead.

## VI. THEORY AND CALCULATION

The eigenfunctions of atoms in an optical lattice are given by the Bloch ansatz as $\phi_q(x) = e^{iqx}u_q(x)$ where $u_q(x) = u_q(x + a)$, $q \in [-k, k]$ is the quasimomentum, and $a$ is the periodicity of the lattice. The field operator for a spin $|\sigma\rangle$, $\Psi^\dagger_\sigma(x)$, and the total field operator for a spin-1/2 particle in a lattice, $\Psi^\dagger_S(x)$, is given as [S5]

$$\Psi^\dagger_\sigma(x) = \sum_{q=-k}^{q=k} \phi^*_q(x) c^\dagger_{q\sigma} \tag{S4}$$

$$\Psi^\dagger_S(x) = (\Psi^\dagger_{|g_1\rangle}(x), \Psi^\dagger_{|g_2\rangle}(x))^T \tag{S5}$$

where $\{c_{q\sigma}, c^\dagger_{q'\sigma'}\} = \delta_{qq'}\delta_{\sigma\sigma'}$.

Before stage IV, all atoms are in $|g_1\rangle$ as the trivial dark state which is represented by the total field operator $\Psi_S^\dagger(x) = \Psi_{|g_1\rangle}^\dagger(x)(1,0)^T$. During stage IV, the adiabatic preparation of the dark-state wavefunction is given as:

$$\Psi_S^\dagger(x) = \Psi_{|g_1\rangle}^\dagger(x)\left(\frac{s\sin(kx)}{\sqrt{s^2\sin^2(kx)+1}}, \frac{1}{\sqrt{s^2\sin^2(kx)+1}}\right)^T \tag{S6}$$

where $s = \frac{1}{\epsilon} = (\Omega_{c1}+\Omega_{c2})/\Omega_p$.

The measurement involves probing the probability density of atoms in $|g_2\rangle$ averaged over the filled ground band of the optical lattice ($|GB\rangle$) using state-selective imaging. The observable that we measure is therefore,

$$\langle g_2|\langle GB|\Psi_S^\dagger(x)\Psi_S(x)|GB\rangle|g_2\rangle = \left(\frac{1}{s^2\sin^2(kx)+1}\right)\sum_{q=-k}^{q=k}\sum_{q'=-k}^{q'=k}\phi_q^*(x)\phi_{q'}(x)\langle GB|c_{q|g_1\rangle}^\dagger c_{q'|g_1\rangle}|GB\rangle \tag{S7}$$

$$= \left(\frac{1}{s^2\sin^2(kx)+1}\right)\sum_{q=-k}^{q=k}|\phi_q(x)|^2 \tag{S8}$$

$$= f(x)\sum_{q=-k}^{q=k}|\phi_q(x)|^2. \tag{S9}$$

Therefore, the measured density distribution within a unit cell is the convolution of the actual density distribution $\sum_{q=-k}^{q=k}|\phi_q(x)|^2$ and the probing function $f(x)$.

We solve for the bandstructure of two types of lattices, Kronig-Penney lattice and sinusoidal ac-Stark shift lattice. Using the Bloch ansatz, the Schrödinger equation can be written as

$$\left(\frac{\hbar^2}{2m}(-i\hbar\partial_x+q)^2+V(x)\right)u_q(x) = \epsilon(q)\,u_q(x). \tag{S10}$$

The Schrödinger equation can be solved numerically by Fourier expansion of $u_q(x)$ into plane waves

$$u_q(x) = \sum_{n=-N}^{N} c_{n,q} e^{inkx}, \tag{S11}$$

where $n \in (0,1,2,...)$ is the band index, and diagonalizing the matrix equation resulting from Eq. S10 in this basis. Similarly, the time dependence of the wavefunction after suddenly changing the lattice can be calculated by solving the time-dependent Schrödinger Eqn. with appropriate initial conditions.

---



\end{document}